\documentclass[summary]{URSIGASS2020}

\usepackage{amssymb}
\usepackage{amsmath}

\usepackage[numbers]{natbib}	

\title{Anchored in Shadows: Tying the Celestial Reference Frame Directly to Black Hole Event Horizons}

\author{T. M. Eubanks}

\affiliation{%
{Space Initiatives Inc\\
Palm Bay, Florida, USA\\
tme@space-initiatives.com\\}
}

\begin{document}

\maketitle

\begin{abstract}

Both the radio International Celestial Reference Frame (ICRF) and the optical 
Gaia Celestial Reference Frame (Gaia-CRF2) are derived from observations of jets produced by
the Super Massive Black Holes (SMBH) powering
active galactic nuclei and quasars. These jets   
are inherently subject to change and will appear
different at different observing frequencies, leading to instabilities and systematic errors in the resulting Celestial Reference Frames (CRFs). Recently, the 
Event Horizon Telescope (EHT), a mm-wave Very Long Baseline Interferometry (VLBI) array, has observed the 40 $\mu$as diameter shadow of the SMBH in M87 at 1.3 mm, showing that the emitting region is smaller than the black-hole shadow.
Use of these SMBH ``emission rings'' (and the associated photon rings) as astrometric references will enable the  resulting CRF to be anchored \textit{directly} in SMBH shadows; the ultimate reference points for any CRF for the forseeable future. A properly equipped space VLBI mission devoted to the observation of SMBH 
event horizons could lead to a two-orders-of-magnitude improvement in the accuracy and stabilty of the ICRF in the relatively near future. 

\end{abstract}

\section{Introduction}

The resolution of VLBI observations at a given frequency are limited by the length of the longest baselines producing significant correlations (i.e., fringes), which has led to numerous proposals for the deployment of VLBI antennas in space.  Of these, three missions to data have acquired VLBI data in space, 
TDRS \cite{Linfield-et-al-1988-b}, HALCA (VSOP) \cite{Hirabayashi-et-al-2000-b} and RadioAstron \cite{Kardashev-et-al-2013-b}, and all three produced fringes. None of these three missions was intended to produce wide-angle astrometric data, due to the inability of large telescopes in free
space to rapidly slew from one source to another, although a limited amount of same-beam phase referencing has been done \cite{Porcas-et-al-2000-a}.

The RadioAstron mission, with a 10 meter space radio telescope used in conjunction with terrestrial VLBI
arrays, found useful VLBI fringes at 1.7, 4.8, and 22 GHz with projected baselines up to 240,000 km, or a
substantial fraction of a lunar distance, producing imaging resolution at the micro arc second level,
and showing that natural radio sources can exhibit brightness temperatures > 10$^{13}$ K 
\cite{Kovalev-et-al-2016-a}. This work clearly
needs to be continued and extended with larger space radio telescopes at even higher frequencies. Further 
impetus to a new space VLBI mission
has been recently provided by the successful imaging of the shadow of M87 by Event Horizon Telescope (EHT)
 \cite{EHT-Collaboration-et-al-2019-a}.
 
\begin{figure}
\centering
\includegraphics[scale=0.65]{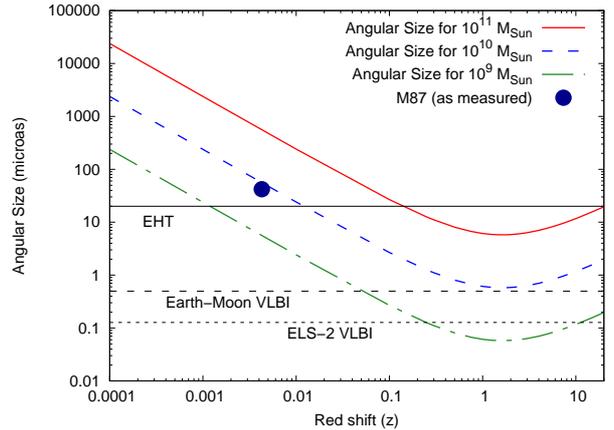}
\caption{
The angular size as a function of red shift and SMBH mass for a reasonable range of AGN masses,
after \cite{Tsupko-et-al-2019-a}, together the resolution of the EHT and 
Earth-Moon and Earth-Sun Lagrange Point 2 (ESL-2) baselines. TON618, the heaviest SMBH known, has a mass of $\sim$6.6 $\times$ 10$^{10}$ solar masses, and so the upper curve, at  10$^{11}$ M$_{\bigodot}$, represents a optimistic upper limit for the sizes of any SMBH shadows in the universe \cite{Vagnozzi-et-al-2020-a}. Note that space baselines will be required to resolve most AGN SMBH shadows out to cosmological distances.}
\label{fig:D_A}
\end{figure}

The EHT is pushing the limits of what is possible with purely terrestrial VLBI, due to the limitations caused atmospheric absorption in the mm and sub-mm bands, which limits both the frequencies used and which limits the EHT to using mostly very dry and high-altitude sites, limiting the (u,v) plane imaging coverage. Even an extension of the EHT to Low Earth Orbit (LEO) would provide a substantial increase in dynamic range and sensitivity over the existing EHT
\cite{Fish-et-al-2020-a}. However, the resolution of the existing EHT, or a LEO augmented EHT, will be limited by the diameter of the Earth and the frequency used. 
The first EHT image of M87 a wavelength of 1.3 mm (230 GHz) and had a resolution of $\sim$20 $\mu$as, close to the fringe spacing of a 1 Earth diameter baseline at that wavelength. 
Figure \ref{fig:D_A} shows that most of the AGN known, which have masses of $\lesssim$ 10$^{10}$ M$_{\bigodot}$ (solar masses) will not be resolved by the current EHT unless they are cosmologically very close to the Earth. 
A substantial improvement in EHT resolution will require the use of VLBI telescopes well removed from the planet. 

VLBI astrometry is based on comparing the VLBI observables (group and phase delay and the change in these quantities) obtained from different sources (see, e.g., \cite{Fey-et-al-2015-a} and references therein). Accurate astrometry requires knowledge or modeling of both the network geometry at the time of observation and also all non-geometric delays or apparent delays in the VLBI system, ideally to within a fraction of the observing wavelength. Non-astrometric delays thus include timing errors, errors in the telescope 
position and media propagation delays. 

All VLBI data is differential (the delay is the time of arrival of a signal at one telescope minus the time of arrival of that signal at another telescope) and effectively all astrometric VLBI data is doubly-differential, the astrometric signal depending on the difference between baseline delay for an observation of one source and an observation of another. 
Any non-geometric delays that do not change between different observations, or whose change can be adequately modeled, can be removed from the astrometric solution using doubly differential data.

In most global (or wide-field) VLBI astrometry, such as the data used to produce the various radio CRFs, observations are not simultaneous, but are as close together in time as possible using fast-slewing telescopes. Also, the geometrical changes in the network, such as Earth rotation changes and continental drift, are typically slow in an Earth-fixed frame and can be estimated from external models or from the VLBI data itself. However, space VLBI telescopes typically have very low slew rates and are of course in orbit, moving rapidly with respect to the Earth, and their orbits frequently have unmodeled or hard to model perturbations. The fast-slewing doubly differential techniques of terrestrial astrometry cannot be applied.

\section{Multi-Beam Multi-Antenna VLBI}
\label{sec:Multi-Beam-VLBI}

Although there is a long history of highly accurate same-beam VLBI \cite{Marcaide-et-al-1984-a}, with standard radio telescopes this technique can only be applied in the unusual cases where two or more sources are within the same telescope beam (i.e., have separations of a few arc-minutes at most). 
The Japanese VERA (VLBI Exploration of Radio Astrometry) project is an array of multi-beam VLBI
antennas used mostly at 22 and 48 GHz for differential for phase referencing extragalactic sources and
astronomical water and silicon oxide masers. 
This system has four receiver bands, at 2, 8, 22, and 43 GHz
and can observe different objects with maximum 2.2 degree separation \cite{Kobayashi-2012-a}, which allows for a reasonable probability of finding a phase reference source near the desired
target. Because the two beams are close together in the sky, and the differential VLBI is done with pairs
of sources with the same clock, with very close paths through the atmosphere, and similar instrumental and
antenna delays, VERA, with an accuracy goal of $\sim$10
$\mu$ as, is capable of highly accurate differential astrometry and shows what be done with narrow-. This dual-beam technique could also be used to extend the sensitivity limit of space VLBI; phase changes observed with a strong source in one beam could be used to correct the phase errors on a much weaker source in the adjacent beam.

In addition to a multiple-beam approach, a multiple-telescope approach seems promising, 
where 2 antennas are mounted together at a fixed or nearly fixed relative angle. 
The space multiple-telescope observing mode would be to point one antenna at a source, while the entire assemblage is rotated about the vector pointing at the first source to point the second antenna at a second source.
This would provide for instantaneous mutual observations, with the same clock and spacecraft position, between two widely separated sources without require any rapid slewing at any time. 

Suppose that the second antenna rotatesat an angle $\theta$ and can cover a total beamwidth of $\epsilon$ radians. 
                For a given primary source, this secondary antenna can cover
\begin{equation}
\mathrm{A} = 2\ \pi\ \epsilon\ \sin{\theta}\ \mathrm{steradians}
\label{eq:coverage}
\end{equation}
of the sky.  For $\theta$ $\sim$ 90 degrees and $\epsilon$ $\sim$ 1 degree, for any given primary source 
the secondary antenna will be able to reach secondary sources in 
about 4 $\times $10$^{-3}$ of the sky, and thus, even given only a few thousand suitable sources, one or more simultaneous source pairs would be generally accessible for any given primary source. Each observation would thus determine the relative positions of 2 widely-separated sources; by repeating this with many pairs of sources a wide-field CRF could be built up over time. 
A multibeam space telescope or telescopes could be used to construct a space-Earth or space-space CRF), providing a substantial increase in both global astrometric accuracy
and differential phase VLBI.

\begin{figure}
\centering
\includegraphics[scale=0.65]{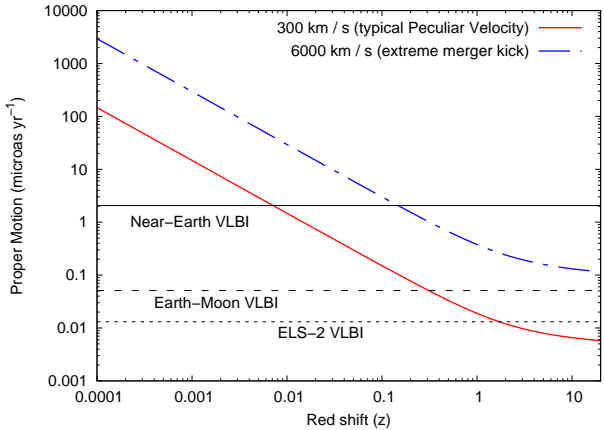}
\caption{Kinematic proper motions for the velocity range expected for non-binary SMBH, together with expected limits from one decade of observing with near-Earth, Earth-Moon and ESL-2 baselines, respectively. For VLBI with ESL-2 baselines, kinematic proper motions should be observable out to cosmological distances.}
\label{fig:Proper-Motions}
\end{figure}

\section{Anchoring on Rings and Shadows}
\label{Rings and Shadows}

The black hole event horizon, of course, basically emits no photons; SMBH shadows are observed through the photons emitted from near, but outside the even horizon.
While SMBH shadows may or may not be suitable ``standard rulers'' for cosmology 
\cite{Tsupko-et-al-2019-a,Vagnozzi-et-al-2020-a}, the photon emission region will 
tied to the mass and spin rate of the host SMBH and will make it possible (in many cases) to parametrically fit for the shadow centroid. Figure \ref{fig:D_A} shows that for many SMBH at cosmological distances the shadow size will be
$\lesssim$1 $\mu$as.

General Relativity predicts that inside the bright emission ring will be ``photon rings,'' narrow arcs representing photons escaping after one or more orbits around the event horizon. 
The photon ring is described as photons escaping from near the critical impact parameter, b = b$_{c}$, where
(for a non-rotating black hole)
\begin{equation}
  \label{eq:critical-b}
  \mathrm{b}_{c} = 3\ \sqrt{3}\ \mathrm{R}_{S}\ \sim\ 5.196\ \mathrm{R}_{S} , 
\end{equation}
and R$_{S}$ is the Schwarzschild radius. 
Photons that are tangent at the critical distance are brought closer to the black hole and are trapped, orbiting an infinite number of times at the bound orbit radius (b = 3 R$_{S}$). Just outside the critical distance unstable orbits exist, where photons can orbit the black hole multiple times and then escape, causing narrow rings in the image plane of the observer. Although these rings can be very bright caustics, Gralla \textit{et al.} \cite{Gralla-et-al-2019-a} conclude that the lensed photon ring and lensing ring (representing different numbers of black hole orbits) are not likely to dominate the total emission, which will mostly come from the emission ring. Although the emission ring can be of variable distance from the center of the SMBH shadow, it will still surround and demark it, providing a suitable reference for accuracy astrometry.

\section{Astrometric Cosmology}
\label{Astrometric-Cosmology}

A multibeam multiantenna VLBI network would enable positioning at the $\mu$as level and below and would, if continued over a suitable duration, also provide information about the angular motions of the SMBH on the sky.
While gravitational lensing events and microlensing will cause transient source motions
\cite{Yano-et-al-2012-a}, most temporal changes at cosmological distances will be secular, appearing as proper motions. 
SMBH shadow proper motions from networks extending out the Moon and beyond will 
provide a sound foundation for the development of  astrometric cosmology \cite{Eubanks-1991-b}
This effort has in fact already begun; the ICRF-3 includes a model for the acceleration caused by the Galaxy, i.e., the aberration rate caused by the rotation of the Solar System Barycenter (SSB) about the center of the Milky Way galaxy, with  the VLBI estimates of the aberration amplitude being in the range of 5.1 to 6.4 $\mu$as yr$^{-1}$ \citep{MacMillan-et-al-2019-a}; this effect is near the accuracy limit of the current VLBI CRF. 

Figure \ref{fig:Proper-Motions} shows that with a network extending to the ESL-2 Lagrange point 
the galactic aberration would be two orders of magnitude above the noise and the expected galactic 
peculiar velocities (typically order 300 km s$^{-1}$) could be detected over much of the recent cosmological past. The ``kicks'' resulting from equal mass SMBH binary mergers can be significantly larger and, although rare, should be easily detectable \cite{Sperhake-et-al-2020-a}.

\section{Conclusions}

A space-Earth or space-space multibeam multiantenna VLBI network would tie the 
Celestial Reference System to SMBH shadows, providing a substantial increase in 
the global astrometric accuracy in the resulting CRF,
and a number of new cosmological  observables.
In a uniform Hubble expansion there are no transverse motions and thus ``all transverse motion is peculiar,'' \cite{Darling-Truebenbach-2018-a}. Any transverse motions detected, whether kinematic \cite{Paine-et-al-2019-a}, or due to gravitational lensing 
\cite{Yano-et-al-2012-a}, or due to gravitational waves \cite{Gwinn-et-al-1997-a} will, provide information about the large scale distribution of matter in the cosmos. The resulting CRF will undoubtedly also find uses in terrestrial and spacecraft navigation and in studies of solar system and galactic dynamics.


\bibliography{IEEEabrv}
\bibliographystyle{ieeetr}

\end{document}